\documentclass[
journal=jpcafh,
manuscript=article]{achemso}

\usepackage[version=3]{mhchem}
\usepackage{booktabs}
\usepackage{threeparttable}
\usepackage{longtable}
\usepackage{xcolor}
\usepackage{textcomp}
\usepackage{soul}
\usepackage{dcolumn}

\title[]{Gas phase detection and rotational spectroscopy of ethynethiol, \ce{HCCSH}}

\author{Kin Long Kelvin Lee}
\affiliation{Harvard-Smithsonian Center for Astrophysics, 60 Garden Street, Cambridge MA 02138, USA}
\email{kinlee@cfa.harvard.edu}
\author{Marie-Aline Martin-Drumel}
\affiliation{Institut des Sciences Mol\'eculaires d'Orsay, CNRS, Univ Paris Sud, Universit\'e Paris-Saclay, Orsay, France}
\author{Valerio Lattanzi}
\affiliation{The Center for Astrochemical Studies, Max-Planck-Institut f\"{u}r extraterrestrische Physik, Garching, Germany}
\author{Brett A. McGuire}
\affiliation{NAASC, National Radio Astronomy Observatory, Charlottesville VA 22903, USA}
\alsoaffiliation{Harvard-Smithsonian Center for Astrophysics, 60 Garden Street, Cambridge MA 02138, USA}
\author{Paola Caselli}
\affiliation{The Center for Astrochemical Studies, Max-Planck-Institut f\"{u}r extraterrestrische Physik, Garching, Germany}
\author{Michael C. McCarthy}
\affiliation{Harvard-Smithsonian Center for Astrophysics, 60 Garden Street, Cambridge MA 02138, USA}

\begin{document}

\begin{abstract}

 We report the gas-phase detection and spectroscopic characterization of ethynethiol (\ce{HCCSH}), a metastable isomer of thioketene (\ce{H2C2S}) using a combination of Fourier-transform microwave and submillimeter-wave spectroscopies. Several $a$-type transitions of the normal species  were initially detected below 40\,GHz using a supersonic expansion-electrical discharge source, and subsequent measurement of higher-frequency, $b$-type lines using double resonance provided accurate predictions  in the submillimeter region.  With these, searches using  a millimeter-wave absorption spectrometer equipped with a radio frequency discharge source were conducted in the range 280 -- 660\,GHz, ultimately yielding nearly 100 transitions up to $^rR_0(36)$ and $^rQ_0(68)$. From the combined data set,  all three rotational constants and  centrifugal distortion terms up to the sextic order were determined to high accuracy, providing a reliable set of frequency predictions to the lower end of the THz band. Isotopic substitution has enabled both a determination of the molecular structure of HCCSH and, by inference, its formation pathway in our nozzle discharge source via the bimolecular radical-radical recombination reaction \ce{SH + C2H}, which is calculated to be highly exothermic (-477\,kJ/mol) using  the HEAT345(Q) thermochemical scheme.

\end{abstract}

\section{Introduction}

From a fundamental perspective, the structure and properties of small organosulfur molecules have long fascinated theorists\cite{siegbahn_theoretical_1983,carsky_ab_1983,bouma_existence_1982,frolov_isomer_2005} and experimentalists alike. \cite{krantz_characterization_1981,krantz_matrix_1974} Much of this interest stems both from the large number of  minima on the potential energy surface that are predicted to exist even when the molecule consists of a relatively small number of atoms, and from the multitude of distinct reaction pathways that might preferentially produce these isomers. Interest in small organosulfur molecules has only intensified in recent years, driven by the discovery that small, hydrogen-deficient sulfur bearing molecules such \ce{C2S} and \ce{C3S} are ubiquitous in cold molecular clouds and other astronomical sources.  In contrast, their hydrogen-terminated counterparts appear conspicuously absent.

\begin{figure}[ht!]
    \centering
    \includegraphics{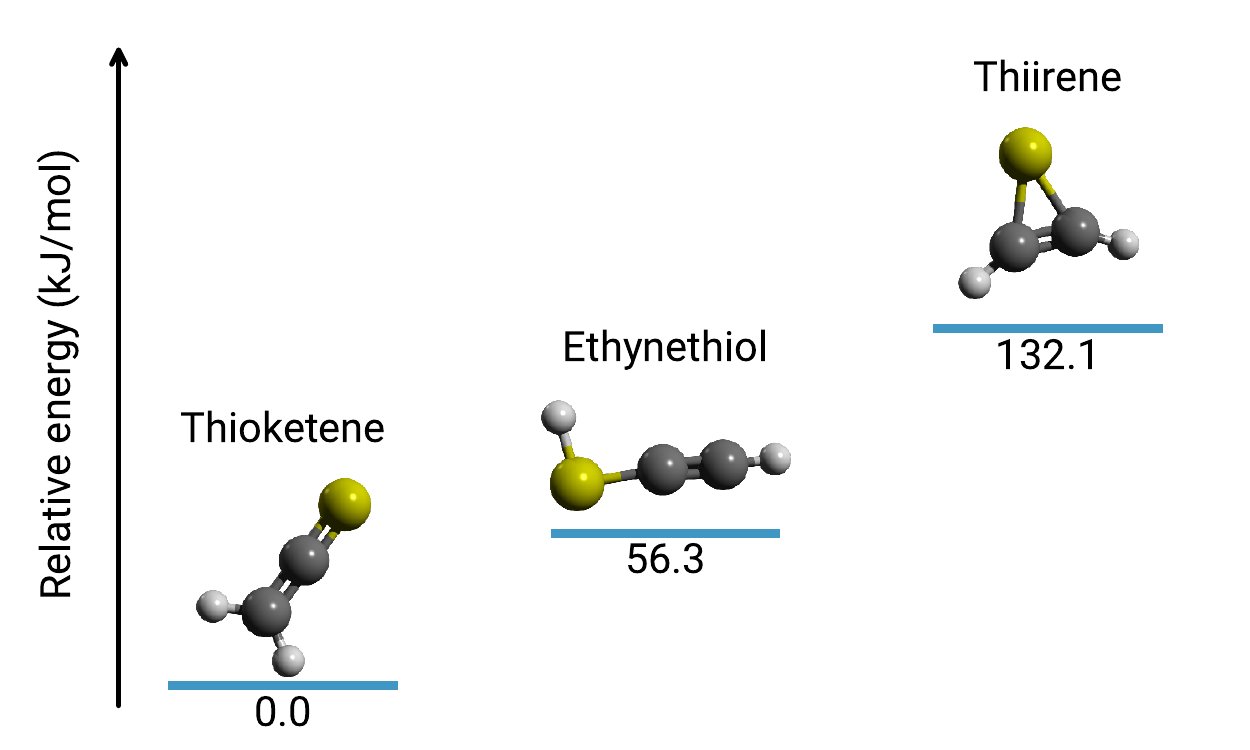}
    \caption{Relative energies of the [\ce{H2},\ce{C2},S] isomers determined with the composite thermochemistry method outlined in the text. Energies are computed at 0\,K, given in kJ/mol, and are relative to the lowest energy form, \ce{H2CCS}.}
    \label{fig:relativeenergetics}
\end{figure}

Organosulfur molecules with the [\ce{H2},\ce{C2},S] formula are one such example. As many as six different stable, singlet isomers have been predicted, with roughly twice that number of triplet variations.\cite{Yamada:2002gm}  Although the stability and relative ordering of the highest energy isomeric arrangements is not fully resolved, there is a general consensus from the electronic structure calculations as to the ordering of the three lowest-energy isomers (Fig.~\ref{fig:relativeenergetics}): thioketene (\ce{H2CCS}), the most stable and the most extensively studied isomer \cite{schaumann_chemistry_1988,kroto_fourier_1985,jarman_high-resolution_1991}; ethynethiol (\ce{HCCSH}), the thioenol form predicted here to lie roughly 60\,kJ/mol higher in energy; and thiirene (c-\ce{H2C2S}), the three-membered heterocycle lying $\sim$135\,kJ/mol above the ground state.\cite{Yamada:2002gm} Due to their high reactivity, the latter two species have remained somewhat enigmatic in the laboratory.  While both are known to form from the decomposition of thiadazoles under ultraviolet irradiation in an argon matrix \cite{krantz_matrix_1974,torres_4n-pi_1980,krantz_characterization_1981,schaumann_chemistry_1988}, the intermediate steps that allow isomerization between the three species remain unclear, despite their importance as potential intermediates in chemical synthesis \cite{braslavsky_gas-phase_1977} and photochemistry \cite{burdzinski_photochemistry_2013,burdzinski_photochemical_2013}. These higher energy isomers may also be of astronomical interest, as their relative abundances in space directly probe the competition between reaction kinetics/dynamics and thermodynamic equilibrium.

Other organosulfur species, specifically the cumulenic C$_n$S\cite{Yamamoto:1987jd,gordon_rotational_2001,mcguire_vibrational_2018}  and H$_2$C$_n$S chains,\cite{gordon_rotational_2002} and the HC$_n$S free radicals, \cite{hirahara_pulsed-discharge-nozzle_1994} have been extensively studied by rotational spectroscopy, largely motivated by their possible astronomical interest. A by-product of these studies has been precise  molecular structures and in some cases detailed information on the electronic distribution. Because small sulfur species, such as \ce{H2S}, only account for a small fraction of the available sulfur in these molecular clouds, it has been widely theorized that the ``missing'' sulfur is sequestered in either the condensed phase or in yet undetected sulfur-bearing molecules. For this reason, astronomical searches for new organosulfur species are often undertaken once accurate rotational line frequencies become available. 

Driven by a desire to better understand the low-lying isomers of small organosulfur species, and to identify possible new transient species that might serve as attractive ``sinks'' for sulfur in molecular clouds,  rotational spectroscopy and high-level theoretical calculations have been undertaken on the [\ce{H2},\ce{C2},S] isomeric system.  Particular emphasis is placed on the second most stable isomer HCCSH, because it plausibly might form directly and efficiently via a recombination reaction involving two well-known astronomical radicals, SH and CCH\cite{Yamada:2002gm}. To fully characterize its rotational spectrum, measurements have been made both at centimeter wavelengths using Fourier transform  (FT) microwave spectroscopy and at submillimeter wavelengths using a direct absorption spectrometer.  To gain insight into the formation pathway of HCCSH in our discharge nozzle source, isotopic investigations have been performed using isotopically-enriched precursors. Complimenting this study are thermochemical calculations using the HEAT protocol have been made to establish the exothermicity of the SH + CCH reaction, and more accurately determine the relative stability of the lower-lying [\ce{H2},\ce{C2},S] isomers. A by-product of the isotopic measurements is a determination of a semi-experimental ($r_{e}^\mathrm{se}$) structure for HCCSH.  This structure is discussed in comparison to  purely equilibrium ones predicted from coupled cluster calculations.  

\section{Experimental and computational methods}
\subsection{Quantum-Chemical Calculations}

Calculations were performed using the CFOUR suite of electronic structure programs.\cite{stanton_cfour_2017} Unless otherwise specified, all-electrons (ae) are correlated in the post-Hartree-Fock (HF) calculations, using coupled cluster methods with single, double, and perturbative triple [CCSD(T)] \cite{stanton_why_1997,bartlett_non-iterative_1990,raghavachari_fifth-order_1989} excitations and correlation consistent basis sets with core-valence basis functions (cc-pCV$X$Z) \cite{woon_gaussian_1993,woon_gaussian_1995} and without (cc-pV$X$Z)\cite{dunning_gaussian_1989} of double ($X$ = D), triple ($X$ = T), and quadruple ($X$ = Q) zeta quality. Accurate predictions of the equilibrium structure of \ce{HCCSH} were obtained by performing geometry optimizations using CCSD(T)/cc-pCV$X$Z ($X$ = D, T, Q), and the various spectroscopic parameters (e.g. equilibrium rotational constants, quartic centrfigual distortion constants) calculated with the cc-pCVQZ geometry. First order vibration-rotation interaction constants required for a semi-experimental molecular structure were calculated under the frozen-core approximation with fc-CCSD(T)/ANO0.

To estimate the reaction thermochemistry, we performed HEAT345(Q) calculations which routinely yield chemical accuracy (${\sim}1$\,kJ/mol).\cite{bomble_high-accuracy_2006,harding_high-accuracy_2008} Since the scheme has been described in previous publications, we only briefly outline it here. Using the CCSD(T)/cc-pVQZ geometry, a series of additive contributions are calculated, including: complete-basis set (CBS) extrapolations of correlation energy using CCSD(T)/aug-cc-pCV$X$Z ($X$ = T, Q, 5); the harmonic zero-point energy (CCSD(T)/cc-pVQZ); extrapolated corrections to the perturbative triple excitations [T - (T)] \cite{scuseria_new_1988}; the diagonal Born-Oppenheimer correction (DBOC) with HF/aug-cc-pVTZ\cite{handy_diagonal_1986}; scalar relativistic corrections [CCSD(T)/aug-cc-pCVTZ]\cite{cowan_approximate_1976,klopper_simple_1997}; and quadruple excitations [fc-CCSDT(Q)/cc-pVDZ]\cite{kallay_approximate_2008}. The contributions to the total composite energy used to calculate the total HEAT345(Q) energy are given in Table \ref{tab:heat}

For completeness, the relative energetics of the lower-lying [\ce{H2},\ce{C2},S] isomers, have also been re-calculated with a lower level of computation sophistication.  For these calculations, the [T - (T)] term has been omitted, and due to linear dependence in the largest basis used for the correlation calculations on c-\ce{H2C2S}, the extrapolated CCSD(T)/CBS calculations utilized the cc-pCV$X$Z ($X$ = T, Q, 5) basis sets instead of their augmented variants are used in the HEAT345(Q) treatment.

\subsection{Fourier-transform microwave spectroscopy}

Experiments at centimeter-wavelengths were conducted first in Cambridge using a FT microwave spectrometer that has been extensively described in previous publications.\cite{mccarthy_microwave_2000} To produce \ce{HCCSH}, acetylene (\ce{HCCH}; 5\,\% in \ce{Ne}) and hydrogen sulfide (\ce{H2S}; 2\,\% in \ce{Ne}) were mixed in-line and further diluted by tenfold with \ce{Ne} at a backing pressure of 2.5\,kTorr. The mixture was introduced along the axis of a microwave cavity  via a discharge nozzle operating at 5\,Hz.  This nozzle source is electrically isolated from the large aluminum cavity mirror on which it is mounted, and, by means of a small hole near the center of the mirror, the gas mixture adiabatically expands into the large vacuum chamber.  By applying a voltage potential between the two cylindrical copper electrodes (1.2\,kV)  in the discharge stack, many collisions with electrons, atoms, and precursor molecules and its fragments occur prior to adiabatic expansion, yielding a rich broth of familiar and exotic molecules. As the gas reaches the beam waist of the cavity, a pulse of resonant microwave radiation polarizes the plasma.  The resulting free-induction decay is detected with a sensitive microwave receiver, the Fourier-transform of which yields the frequency spectrum. Powerful and flexible in-house software is used to control, optimize, and acquire data in the 5--40\,GHz frequency range of this spectrometer. For isotopic measurements, isotopically-enriched precursors gases such as \ce{D2S}, DCCD, H$^{13}$C$^{13}$CH, etc., were used instead of the normal sample at the same level of dilution.

Once candidate $a$-type lines of HCCSH were identified, subsequent high-frequency, $b$-type transitions were sought using double resonance, since these lines are predicted to  lie well above the frequency ceiling of the microwave cavity.  In this type of experiment, the cavity spectrometer is tuned to the frequency of a low-$J$, $a$-type line, and radiation generated from an active multiplier chain in combination with a second synthesizer is aligned to intersect the beam waste of the cavity.  The frequency of this second radiation source is sequentially stepped in  small intervals (typically 0.1\,MHz) so as to cover the frequency range predicted for the transition.  A significant decrease in line intensity is normally observed when the two rotational transitions share a common upper or lower rotational level owing to loss of coherence. Although many hundreds or even thousands of steps may be required to detect these lines, wide frequency sweeps -- covering a GHz or more -- are readily performed under computer control. The frequency precision attained by the cavity measurements is on the order of 2\,kHz. Due to broader lineshapes and lower signal-to-noise ratio, frequency precision of the order of (${\sim}$30 -- 50\,kHz) is achieved for double resonance experiments.

\subsection{Millimeter-wave direct absorption spectroscopy}
The submillimeter experiments were performed in Orsay in the 280--660\,GHz region using a direct absorption spectrometer \cite{Pirali2017}. Briefly, the radiation from a post-amplified synthesizer (7.8--12.2\,GHz) drives a commercial frequency multiplier source (Virginia Diodes, Inc.). An off-axis parabolic mirror collimates the radiation from the multiplication chain into a 1.2\,m long single path Pyrex absorption flow cell equipped with teflon windows, and the output radiation is focused onto a liquid-helium cooled Si-bolometer detector. 
The flow cell is equipped with a radio frequency resonator driven by a generator that can provide as much as 100\,W\cite{Martin-Drumel2011}. In the present experiment, the gas flow was maintained by a mechanical pump (pumping speed 28 m$^3$/h). 
Typical experimental conditions consisted of a 10\,W radio frequency discharge; a flowing mixture of  \ce{HCCH} and \ce{H2S} in a 1:1 pressure ratio at a total pressure of 50\,\textmu bar; a 30--50\,kHz frequency step size; a 49\,kHz frequency modulation (resulting in a second derivative line shape of the recorded transitions) with a modulation depth of 450 kHz; and a 200\,ms time constant. Under these conditions, rest frequencies were determined to an accuracy of 50\,kHz.

\section{Results and discussion}

\subsection{On the reliability of the \textit{ab initio} structure determination and dipole moment}

\begin{figure}[ht!]
    \centering
    \includegraphics[width=0.4\textwidth]{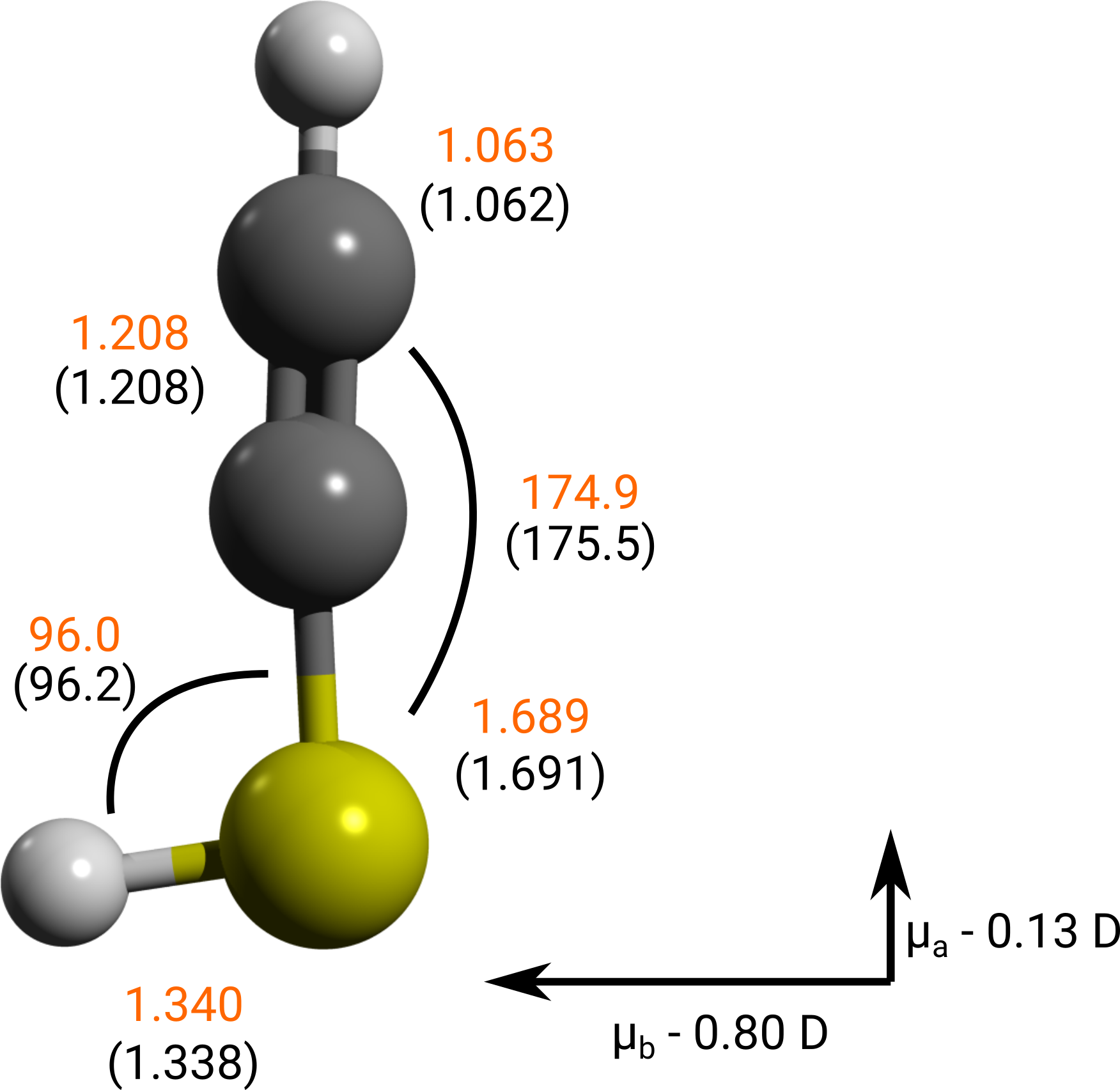}
    \caption{Equilibrium structure of \ce{HCCSH}. Bond lengths are given in \AA, angles in degrees. Values in parentheses (and in black) are obtained at the ae-CCSD(T)/cc-pCVQZ level, while the orange parameters correspond to the best-fit semi-experimental parameters (Table \ref{tab:bonding}). Dipole moments (in Debye) are calculated using the CCSD(T)/cc-pCVQZ structure at the same level of theory.}
    \label{fig:structure}
\end{figure}

The rotational constants and projections of the dipole moment derived with each basis set are reported in Table \ref{tab:abinitio}, while the structural parameters obtained at the largest basis (cc-pCVQZ) is shown in Figure \ref{fig:structure}. Regardless of the basis set, the calculations consistently predict that \ce{HCCSH} is a near-prolate asymmetric top (asymmetry parameter, $\kappa=-0.999$), with a heavy atom linear backbone nearly coincident with the $a$ inertial axis.  Since the S--H bond is close to perpendicular with respect to the heavy backbone, the $a$ and $b$ components of the electric dipole moment are both non-zero, and the $A$ rotational constant is significantly larger than $B$ and $C$ (${\sim}290$\,GHz compared to roughly 5 GHz). 

By performing systematic geometry optimizations, it is possible to assess the convergence of the rotational constants and dipole moments with increasing size of correlation consistent basis sets. As summarized in Table \ref{tab:abinitio}, the dipole moments converge quickly with the size of the basis set although the relative change is different for $\mu_a$ and $\mu_b$: the former increases with increasing basis, while the latter decreases. The magnitudes of the two moments can be rationalized in the following way: $\mu_b$ is dominated simply by the polarity of the S--H bond since all the other atoms in \ce{HCCSH} lie very close to the $a$ inertial axis. For this reason it is perhaps not surprisingly that the value of $\mu_b$ calculated here ($\mu_a$=0.80\,D) is very similar to that measured for free SH ($\mu$=0.758\,D; Ref.~\citenum{Meerts_1974_l45}).  The magnitude of $\mu_a$ is more subtle. The CCS radical is calculated to be highly polar (2.8\,D; Ref.~\citenum{Murakami_1990}) but, in contrast to \ce{HCCSH}, possesses a cumulenic :C=C=S: like-structure with formally a lone pair on both the terminal carbon and sulfur atoms.  Addition of an H atom to the terminal C  fundamentally alters this bonding, imparting an acetylenic-like conjugation to the heavy atom backbone.  The absence of a lone pair on the C atom substantially diminishes the polarity relative to free \ce{CCS}, with our calculations suggesting at the highest levels of theory that $\mu_a$ is vanishingly small, comparable to that of CO. 

With respect to the rotational constants, larger basis sets tend to lead to bond length contraction, which in turn decreases the three moments of inertia, and increases the magnitude of these constants, as  Table \ref{tab:abinitio} illustrates. The fractional increase in the rotational constants with respect to the size of the basis set is similar for all three constants (of the order of 2.5\,\% change between cc-pCVDZ and cc-pCVTZ and 0.5\,\% between cc-pCVTZ and cc-pCVQZ). However, owing to the much larger value of the $A$ constant, even small fractional differences still correspond to large changes in its magnitude.  For example, $A_e$ changes by ${\sim}$600\,MHz between cc-pCVTZ and cc-pCVQZ, while $B_e$ and $C_e$  only differ about 30\,MHz with respect to the same two basis sets.  Not withstanding the small $\mu_a$ for \ce{HCCSH}, quite reliable estimates of its low-frequency $a$-type lines (which scale as integer multiples of $B +C$) can be made, but much larger and therefore time-consuming searches are required to detect its high-frequency $b$-type lines (which are roughly $A+C$). Despite the small $\mu_a$, a key advantage of FT microwave spectroscopy is that line intensities scale as $\mu$ as opposed to $\mu^2$ in conventional absorption or emission spectroscopy, allowing weak polar species to be routinely detected by this technique.

\begin{table}[ht!]
    \centering
    \caption{Equilibrium rotational constants (in MHz) of \ce{HCCSH} following optimization at the ae-CCSD(T)/cc-pCV$X$Z ($X$ = D, T, Q) level. Dipole moments are evaluated using the CCSD(T)/cc-pCVQZ equilibrium geometry, and their absolute magnitudes are given in Debye. The last two columns indicate the fractional changes of the rotational constants as the size of the basis set increases.}
    \begin{tabular}{lrrr|rr} \toprule
    ~ &       cc-pCVDZ &       cc-pCVTZ &       cc-pCVQZ & D $\rightarrow$ T & T $\rightarrow$ Q \\
    \midrule
    $A_e$ &  285850 &  292903 &  293556  & 2.4\% & 0.2\%\\
    $B_e$ &    5372 &    5513 &    5545 & 2.6\% & 0.6\% \\
    $C_e$ &    5273 &    5411 &    5443 & 2.6\% & 0.6\%\\
    \midrule
    $\mu_a$ & 0.03 & 0.09 & 0.13 \\
    $\mu_b$ & 0.90 & 0.82 & 0.80 \\
    \bottomrule
    \end{tabular}
    \label{tab:abinitio}
\end{table}

\subsection{Laboratory investigation of the \ce{HCCSH} rotational spectrum}

From the \textit{ab initio} structure, the three lowest \textit{a}-type transitions; $J = 1_{01} - 0_{00}$ ($J=1-0$), $2_{02} - 1_{01}$ ($2-1$), and $3_{03} - 2_{02}$ ($3 - 2$), predicted at roughly $11$, $22$, and $33$\,GHz, respectively, lie well within the frequency range of our FT microwave spectrometer.  A search for the fundamental $a$-type transition was undertaken first, and soon after yielded an unidentified line within a few MHz of our best estimate (at 10,985\,MHz compared to the predicted value of 10,988\,MHz from the equilibrium constants, see Table \ref{tab:abinitio}). Subsequent screening tests have established that i) this spectral feature requires an electrical discharge; ii) its line intensity is insensitive to the presence of an external magnetic field; and iii) it requires both precursor gases, \ce{H2S} and HCCH. Furthermore, the line intensity is maximized at high microwave powers, implying the carrier has a small dipole, which we roughly estimate to be $\leq0.5$\,D. Surveys for the next two rotational transitions each resulted in an  unidentified line, systematically  offset in frequency from the rigid rotor prediction --- i.e.~neglecting centrifugal distortion. These two new lines behave in the same manner as the line at 10,985\,MHz, and the frequencies of all three are well reproduced (RMS of 0.2\,kHz) using a linear molecule Hamiltonian with one free parameter $(B+C)/2$.  The best-fit value is within 2\,\% of the \textit{ab initio} prediction for $(B+C)/2$.

\begin{figure}[ht!]
  \includegraphics[width=0.4\textwidth]{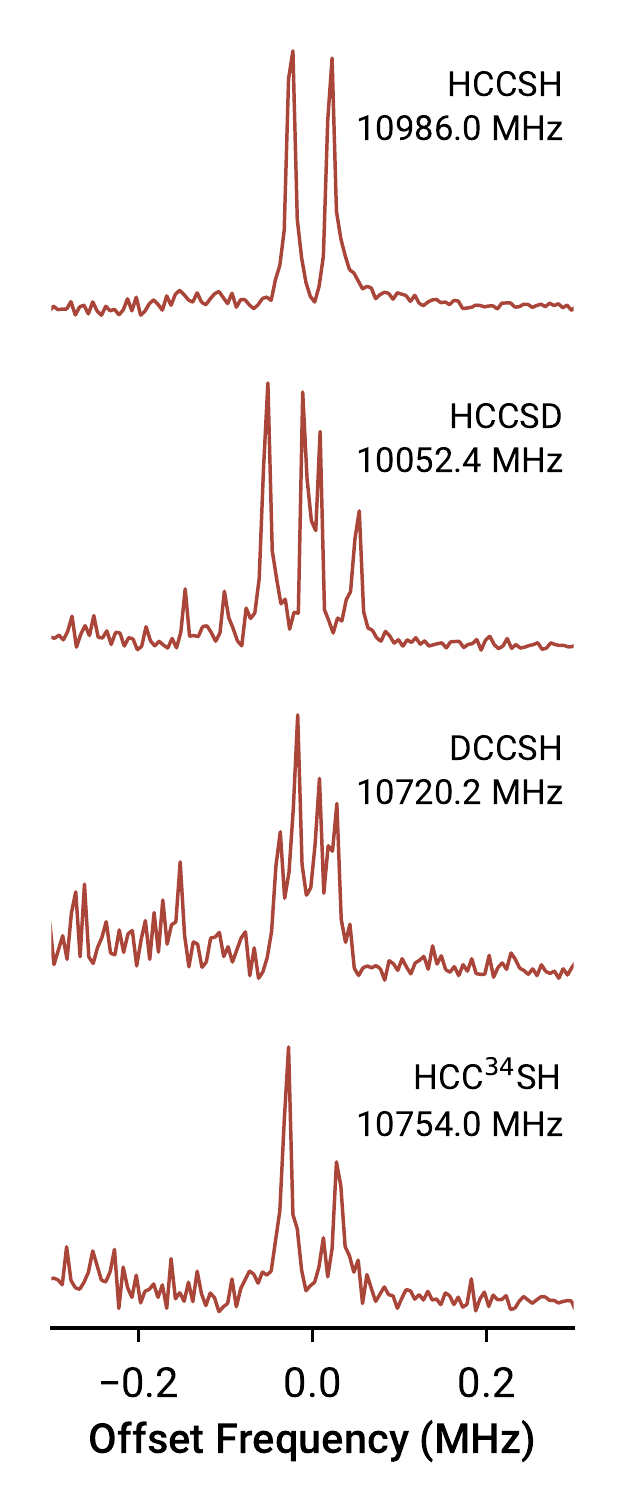}
  \caption{Representative spectra of the fundamental $J=1-0$ transition for \ce{HCCSH}, \ce{HCCSD}, \ce{DCCSH}, and \ce{HCC$^{34}$SH}. The rest frequency (used to calculate the offset frequency in the abscissa) of each transition is indicated. In addition to Doppler splitting, partially- or well-resolved hyperfine splitting is apparent in the spectra of the two deuterated species.}
  \label{fig:sulfur-isotopes}
\end{figure}

To identify the carrier of the new molecule, ostensibly \ce{HCCSH},  with greater confidence, searches for its isotopologues, the most intense based on natural abundance is $^{34}$S (4.21\,\%), were performed. Soon afterwards, a weak line corresponding to the $J=1-0$ transition of \ce{HCC$^{34}$SH} was found within 1\,MHz of the frequency predicted by scaling the theoretical rotational constants (see the supplementary material), as were the $J=2-1$ and $3-2$ lines at higher frequency. In addition, by substituting \ce{HCCH} with \ce{DCCD}, and subsequently \ce{H2S} with \ce{D2S}, lines of both \ce{DCCSH} and \ce{HCCSD} were also observed very close in frequency to the predictions.  In the spectra of both, the presence of partially or well-resolved hyperfine-splitting structure arising from the deuteron (Fig.~\ref{fig:sulfur-isotopes}) lends further support that the carrier of the new lines is \ce{HCCSH} or one of its isotopologues, and no other molecule. Finally, by using a statistical mixture of \ce{HCCH}, \ce{H^{13}CCH}, and \ce{H^{13}C^{13}CH} as well as pure \ce{H^{13}C^{13}CH}, all possible \ce{^{13}C} variants were also detected, namely  \ce{H^{13}CCSH}, \ce{HC^{13}CSH}, and \ce{H^{13}C^{13}CSH}. The transition frequencies for all the \ce{HCCSH} isotopologues measured in the course of the present work are summarized in the supplementary material.

\begin{figure}
    \centering
    \includegraphics[width=\textwidth]{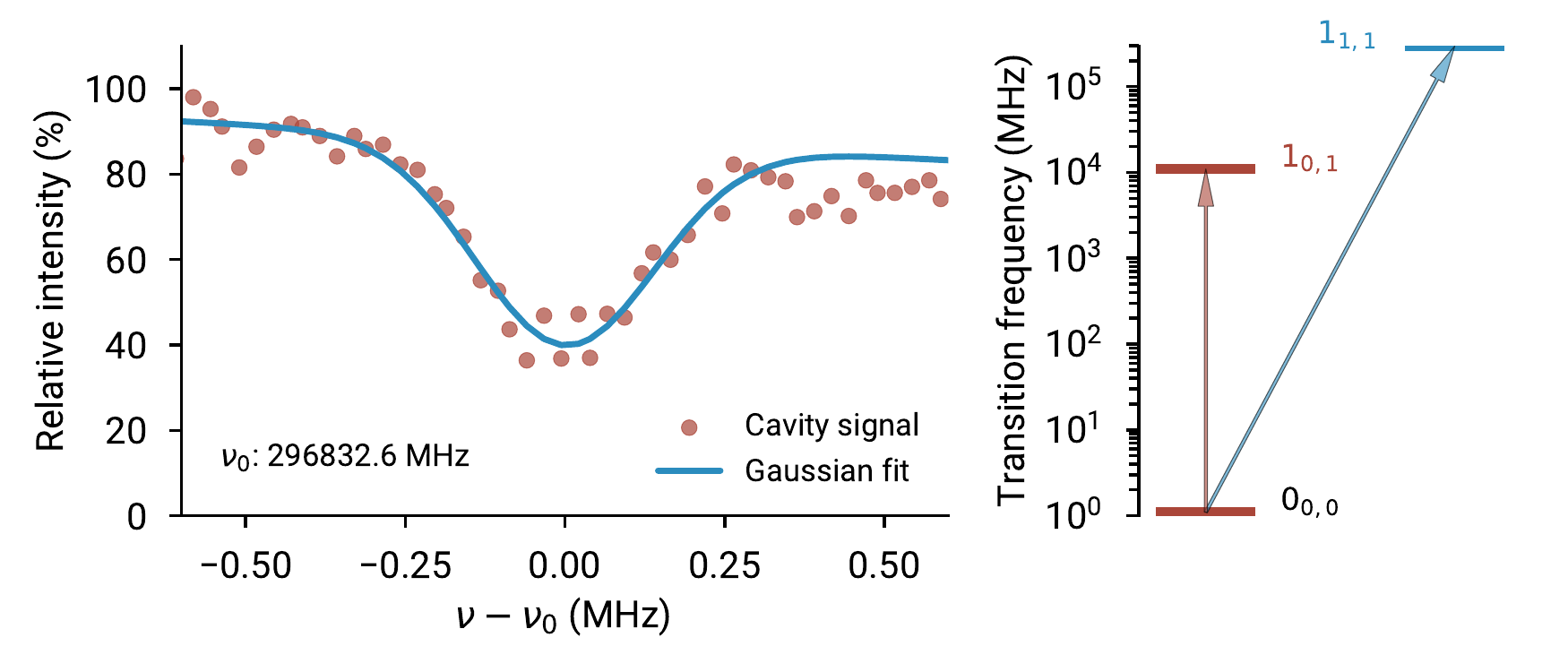}
    \caption{A portion of the double resonance survey for the fundamental \textit{b}-type transition ($J=1_{11} - 0_{00}$, blue) of HCCSH, performed by monitoring the intensity of its \textit{a}-type line at 10,985\,MHz ($J=1_{01} - 0_{00}$, red). The full survey was acquired with 50\,kHz steps, and required 72 hours to complete. A clear (${\sim}$50\,\%) depletion of the \textit{a}-type signal was observed, and fit with a Gaussian and linear baseline. The full-width half-maximum linewidth of the Gaussian was 330\,kHz.
    \label{fig:drsurvey}}
\end{figure}

Although we can be confident from the centimeter-wave measurements that \ce{HCCSH} has been identified, this data alone provides only fragmentary information on the underlying rotational constants, and provides little predictive power of the higher-frequency spectrum, particularly its intense $b$-type transitions. For these reasons, a search for low-$J$, $b$-type lines was initiated using double resonance.  Although the predicted transition frequency (${\sim}298$\,GHz) of the fundamental $b$-type line ($1_{1,1}$ --- $0_{0,0}$) lies far above the operating range in the FT microwave spectrometer, the lower level in this transition ($0_{0,0}$) is also the lower level of the fundamental $a$-type transition.  Hence, by monitoring the $a$-type line  as the frequency of the  
millimeter-wave radiation is varied, it should be possible to detect a depletion of its intensity, provided enough millimeter-wave power is available to saturate the high-frequency transition.  Although initial surveys close to the predicted frequency of 298\,GHz, were unsuccessful, eventually a clear depletion was detected nearly 2\,GHz lower in frequency, as depicted in Figure \ref{fig:drsurvey}, a difference which reflects the large relative uncertainty inherent in the prediction of the $A$ constant. Consequently, the double resonance survey required 72 hours to complete. Following the initial discovery, two other $b$-type transitions were measured in the same fashion, as were an analogous set of lines for \ce{HCCSD}. By combining the millimeter-wave and centimeter-wave measurements, it was possible to determine preliminary values for all three rotational constants.

\begin{figure}
    \centering
    \includegraphics[width=\textwidth]{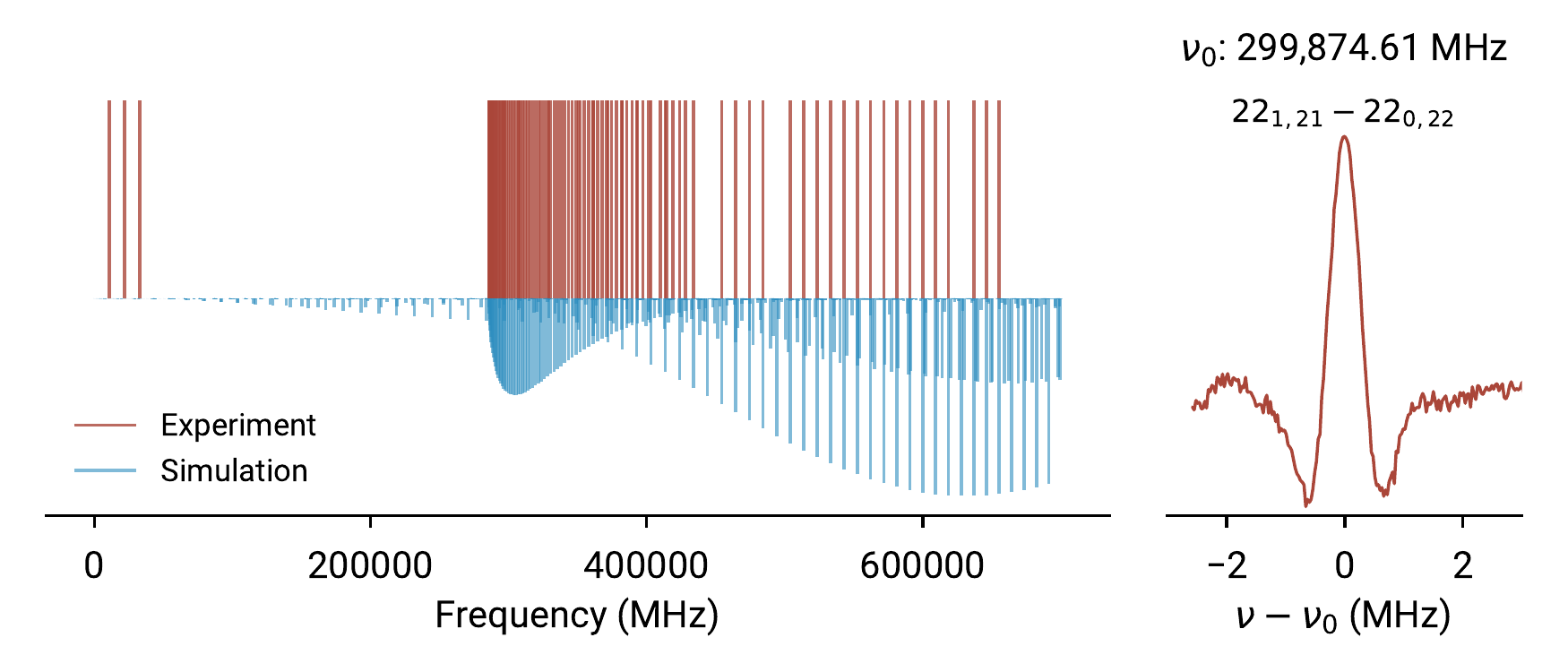}
    \caption{\textit{Left:} Stick representation of the experimental observations (top) and a 300\,K simulation (bottom) of the rotational spectrum of \ce{HCCSH}. The simulated spectrum was calculated using the experimental constants presented in Table \ref{tab:rotconstants}. \textit{Right:} Representative trace of the millimeter-wave absorption spectrum; the abscissa indicates the center frequency of the transition.}
    \label{fig:mmw}
\end{figure}

From frequency predictions derived from these best-fit constants, the spectroscopy of \ce{HCCSH} has been extended into the millimeter and submillimeter regime using a standard absorption spectrometer. Because many levels are thermally populated in the room temperature radio-frequency discharge source, it was straightforward to measure many $b$-type lines in this spectral region (Figure \ref{fig:mmw}). Under our experimental conditions, strong transitions ---mainly $K_a'' = 0$ $b$-type lines--- of \ce{HCCSH} at 300\,K fall within the range of the spectrometer. In total, an additional 93 lines up to the $^rR_0(36)$ and $^rQ_0(67)$ were assigned. We note that frequency coverage was limited by the output of the active multiplier chain units, rather than the temperature of our sample.

\subsection{Determination of the spectroscopic parameters}

Spectroscopic parameters were determined using the SPFIT/SPCAT suite of programs \cite{Pickett:1991cv}. Owing to the highly prolate character of HCCSH ($\kappa=-0.999$) and the similar magnitudes  of $B$ and $C$, a Watson-$S$ Hamiltonian in the $\mathrm{I^r}$ representation was employed. For the main isotopic species, a total of 9 free parameters were required to reproduce the available data to an RMS of 46\,kHz (Table \ref{tab:rotconstants}). In addition to the three rotational constants, three of the five quartic centrifugal distortion terms were varied, as was one sextic ($h_1$) off-diagonal term, while $D_K$ and $d_2$ were constrained to their \textit{ab initio} values. 
As indicated in Table \ref{tab:rotconstants}, the \textit{ab initio} equilibrium constants are in extremely good agreement with the experimentally-derived ground state constants, within 1\% or less. When comparison is possible, the experimental centrifugal distortion constants are also in good agreement with the \textit{ab initio} values. The experimentally-derived inertial defect (0.093\,amu\,\AA$^2$ for HCCSH and 0.0129\,amu\,\AA$^2$ for HCCSD) is consistent with a planar geometry, as  expected from the \textit{ab initio} calculations.

\begin{table}[ht!]
  \centering
  \caption{Spectroscopic parameters for \ce{HCCSH} obtained by fitting the rotational transition frequencies to a $S$-reduced Hamiltonian. For comparison, the ae-CCSD(T)/cc-pCVQZ equilibrium rotational constants and quartic centrifugal distortion constants are provided together with the error on these predictions expressed in percentage of the experimental value [$\delta=(exp.-calc.)/calc*100$]. Parameters are given in units of MHz; values in parentheses correspond to $1\sigma$ uncertainty. 
  }
  \label{tab:rotconstants}
  \begin{threeparttable}
  \newcolumntype{.}{D{.}{.}{-1}}
  \begin{tabular}{l ...}
    Parameter & \multicolumn{1}{c}{Experimental fit} & \multicolumn{1}{c}{\textit{Ab initio}} & \multicolumn{1}{c}{$\delta$ /\%}\\
    \toprule
    $A$ & 291414.3934\,(110) & 293555.696 & -0.73\\
    $B$ & 5547.541331\,(308) & 5545.43900 & 0.04\\
    $C$ & 5438.444587\,(246) & 5442.62463 & -0.08\\
    $D_J \times 10^{3}$ & 1.37975\,(69) & 1.3290 & 3.82\\
    $D_{JK}$ & 0.136857\,(55) & 0.13925 & -1.72 \\
    $D_K$ & 19.8358\tnote{a} & 19.8358 \\
    $d_1 \times 10^{6}$ & -0.0292759\,(160) & -0.024834 & 1.79\\
    $d_2 \times 10^{6}$ & -3.20926\tnote{a} & -3.20926 & ~ \\
    $h_1 \times 10^{9}$ & 0.11969\,(272) & ~ & ~ \\
    \bottomrule
  \end{tabular}
  \begin{tablenotes}
    \item[a] Value fixed to the CCSD(T)/cc-pCVQZ value
  \end{tablenotes}
  \end{threeparttable}
\end{table}

The best-fit spectroscopic constants for all six rare isotopic species are reported in the supplementary material. Owing to the small number of measured rotational transitions for most of these species, a complete spectroscopic analysis was not feasible. In these cases, some rotational constants were fixed to the corresponding \textit{ab initio} value scaled by the ratio between the experimental and \textit{ab initio} values for the same constant of the main isotopic species, while the centrifugal distortion parameters were fixed to the purely \textit{ab initio} values.

\subsection{Formation Chemistry}

To establish if there is a clear and dominant mechanism responsible for \ce{HCCSH} in our discharge nozzle, a   systematic series of isotopic labelling studies were performed using our FT microwave spectrometer. The strong preferential production of \ce{HCCSD} when \ce{D2S} was used in place of \ce{H2S} as a precursor gas, and the absence of detectable quantities of \ce{HCCSH} when \ce{CS2} is used as an alternative source of sulfur, provide very strong evidence that the \ce{SH} radical plays a central role in molecule formation.

\begin{figure}
    \centering
    \includegraphics[width=\textwidth]{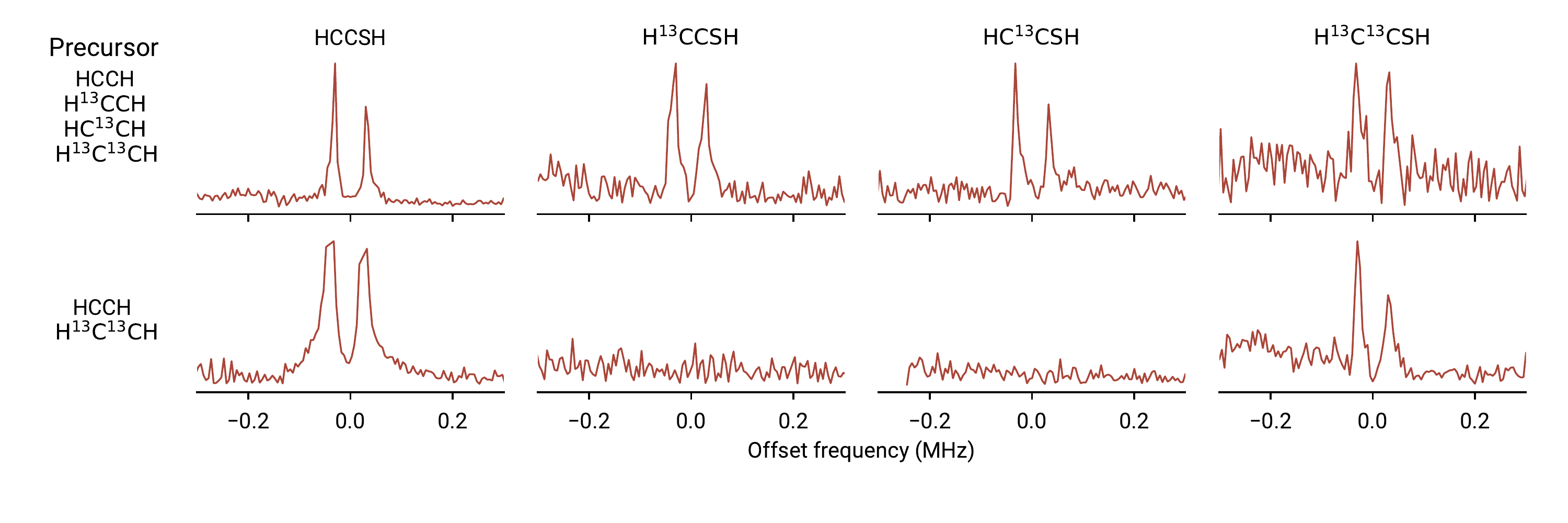}
    \caption{Carbon-13 enriched assays of \ce{HCCSH} and its isotopologues using two hydrocarbon samples. Each spectrum displays the $J=1-0$ transition of the four $^{13}$C species of \ce{HCCSH}, displayed as an offset with respect to its rest frequency as listed in the SI. The top row indicates the results obtained with a statistical mixture of carbon-13 enriched \ce{HCCH}, while the bottom row is the same set of measurements, but with a mixture of pure \ce{HCCH} and \ce{H^{13}C^{13}CH}. Each scan was accumulated for one minute at a collection rate of 5\,Hz.}
    \label{fig:hcch-chem}
\end{figure}

To explore the formation mechanism in greater detail, two different sources of $^{13}$C were also tested: a statistical mixture of $^{12}$C/$^{13}$C acetylene (i.e.~roughly 25\% \ce{HCCH}, 50\,\% \ce{H^{13}CCH}, and 25\,\% \ce{H^{13}C^{13}CH}), and roughly equal mixtures of normal acetylene (\ce{HCCH}) and $^{13}$C-acetylene (\ce{H^{13}C^{13}CH}). Figure \ref{fig:hcch-chem} shows the assay matrix, in which the fundamental rotational transition of each $^{12}$C/$^{13}$C species was sequentially measured using the two hydrocarbon mixtures.  As illustrated in this Figure, lines of all four $^{13}$C species were readily detected with the statistical sample.  With equal mixtures of \ce{HCCH} and \ce{H^{13}C^{13}CH}, however, only \ce{HCCSH} and \ce{H^{13}C^{13}CSH} were readily observed as discharge products, implying that there is little or no scrambling of the carbon atoms and that the \ce{C2} unit remains intact during molecule formation. In combination with the preferential formation of \ce{DCCSH}  using \ce{DCCD} as a precursor, we conclude with a high degree of confidence that the reaction most likely involves the \ce{C2H} radical.  Taken together, all of the available isotopic data is consistent with a simple and direct pathway to form HCCSH in our discharge: homolytic cleavage of the H--S bond of \ce{H2S} and the H--C bond in acetylene, followed by radical-radical recombination. This conclusion is supported by HEAT345(Q) calculations, with the individual contributions organized in Table \ref{tab:heat}. The total HEAT energies are used to calculate the standard 0\,K reaction enthalpy for \ce{SH + CCH -> HCCSH}, which is determined to be -477.2\,kJ/mol with a nominal statistical uncertainty of $\pm1$\,kJ/mol.\cite{karton_computational_2016} This value is in qualitative agreement with the lower level estimations by \citet{Yamada:2002gm}, who used B3LYP geometries and zero-point energies combined with CCSD(T)/aug-cc-pVTZ energies to obtain a value of -436\,kJ/mol.

\begin{table}[ht!]
    \centering
    \caption{Breakdown of the contributions to the HEAT345(Q) energy for \ce{C2H}, \ce{SH}, and \ce{HCCSH}. Energies are given in Hartrees.\label{tab:heat}}
    \begin{threeparttable}[]
        \begin{tabular}{l r r r}
             Contribution & \multicolumn{1}{c}{\ce{HCCSH}} & \multicolumn{1}{c}{\ce{SH}} & \multicolumn{1}{c}{\ce{C2H}} \\
             \toprule
             SCF/CBS     & -474.410257 & -398.110854 & -76.183645 \\
             CCSD(T)/CBS &   -1.132275 &   -0.639181 &  -0.430304 \\
             ZPE         &    0.027522 &    0.006224 &   0.014876 \\
             MVD         &   -1.143031 &   -1.111314 &  -0.032137 \\
             HLC-(T)     &   -0.582541 &   -0.204518 &  -0.315919 \\
             HLC-T       &   -0.582475 &   -0.205373 &  -0.317067 \\
             CCSDT(Q)\tnote{a}    &   -0.000929 &   -0.000120 &  -0.000611 \\
             DBOC        &    0.009462 &    0.005953 &   0.004008 \\
             T - (T)\tnote{b}     &    0.000066 &   -0.000855 &  -0.001148 \\
             \midrule
             HEAT345(Q)  & -476.649442 & -399.850147 & -76.628962 \\
             \bottomrule
        \end{tabular}
        \begin{tablenotes}
            \item[a] Correlation contribution from fc-CCSDT(Q)/cc-pVDZ
            \item[b] Difference in the extrapolated fc-CCSD(T) and fc-CCSDT energy
        \end{tablenotes}
    \end{threeparttable}
\end{table}

\subsection{The molecular structure of HCCSH}

Because a large number of isotopic species have been observed in the present work, it is possible to derive both experimental ($r_0$) and semi-experimental ($r_e^\mathrm{se}$) structures for HCCSH. In either structural determination, the six unique structural parameters, the four bond lengths, and the two bond angles (CCS and CSH) as depicted in Fig.~\ref{fig:structure}, were optimized using a standard non-linear least-squares minimization procedure\cite{kisiel:58} to reproduce to the nine moments of inertia of all seven isotopic species. A planar structure was assumed. For the $r_0$ structure, $B$ and $C$ for both HCCSH and HCCSD (Table \ref{tab:rotconstants}) along with $B+C$ for the remaining isotopic species are used, while for the $r_e^\mathrm{se}$ structure, these constants are first corrected for zero-point vibrational motion, as calculated theoretically using second-order vibrational perturbation theory (VPT2) \cite{stanton_analytic_2000,schneider_anharmonic_1989}. The corrected rotational constants are derived using the equation: $B_e \approx B_0 + \alpha_0$, 
where $\alpha_0$ are vibration-rotation interaction constants to first order, which are calculated at the fc-CCSD(T)/ANO0 level of theory (See supplementary material).  For normal HCCSH, correction of the three rotational constants reduces the inertial defect from 0.093 to 0.011\,amu~\AA$^2$, suggesting both the electronic structure method  and second-order vibrational perturbation theory accurately treat the vibrational structure.

Table~\ref{tab:bonding} summarizes the best-fit structural parameters in comparison the purely \textit{ab initio} geometry ($r_e^\mathrm{theo}$). Although all seven parameters were determined in the $r_e^\mathrm{se}$ structure, it is not possible to determine the small predicted departure of the CCS angle from linearity in the $r_0$ structure, so this angle was simply fixed at the theoretical value. This difference aside, the two experimental structures are very similar: the heavy atom bond lengths are not statistically different, but as might be expected when vibrational corrections are included, both the S--H and C--H bonds contract slightly from the $r_0$ to the $r_e^\mathrm{se}$ structure.

The $r_e^\mathrm{se}$ remarkably well reproduces the available isotopic data: all nine constants are reproduced to better than 0.06\,MHz, resulting in statistical uncertainties in the sub-m\AA~range for the bond lengths. Furthermore, this structure and the equilibrium structure calculated with theory are in near perfect agreement: differences among the four bond lengths amount to no more than 2\,m\AA, while the differences between the two angles are at most 0.6$^{\circ}$. 

\begin{table}[ht!]
    \centering
    \caption{The experimental ($r_0$) and semi-experimental ($r_e^\mathrm{se}$) structures, in comparison to the equilibrium structure obtained at the ae-CCSD(T)/cc-pCVQZ level ($r_e^\mathrm{theory}$). \label{tab:bonding}}
\begin{threeparttable}[]
  \newcolumntype{.}{D{.}{.}{-1}}
    \begin{tabular}{l ...}
        Parameter\tnote{a} & \multicolumn{1}{c}{$r_0$} & \multicolumn{1}{c}{$r_e^\mathrm{se}$} & \multicolumn{1}{c}{$r_e^\mathrm{theory}$} \\
        \toprule
        $r_\mathrm{HC}$ & 1.056(1) & 1.0627(1) & 1.062 \\
        $r_\mathrm{CC}$ & 1.209(3) & 1.2082(3) & 1.208 \\
        $r_\mathrm{CS}$ & 1.691(2) & 1.6892(2) & 1.691 \\
        $r_\mathrm{SH}$ & 1.366(2) & 1.3403(4) & 1.338 \\
        $\theta_\mathrm{CCS}$ & 175.5\tnote{b} & 174.93(18) & 175.5 \\
        $\theta_\mathrm{CSH}$ & 95.33(15) & 96.04(5) & 96.2 \\
       \bottomrule
    \end{tabular}
           \begin{tablenotes}
            \item[a] Bond lengths in Angstroms, bond angles in degrees. Values in parentheses are formal $1\sigma$ statistical uncertainties.
            \item[b] Fixed to the \textit{ab initio} equilibrium value.
        \end{tablenotes}
    \end{threeparttable}
\end{table}

\subsection{Stability and relative abundances of the [\ce{H2},\ce{C2},S] isomers}

The re-computed relative energies of the three lowest energy [\ce{H2},\ce{C2},S] isomers are shown in Figure \ref{fig:relativeenergetics}. The values derived here are in qualitative agreement with those obtained with the latest MP2/6-311G(2d,p) and B3LYP/6-311G(2d,p) calculations by \citet{frolov_isomer_2005}, and other calculations\cite{Yamada:2002gm}.  In all cases, the energy ordering of the three isomers is identical, however, there are quantitative differences in the relative energetics. For \ce{HCCSH}, the B3LYP determination (76.2\,kJ/mol) is much larger than that derived from the MP2 calculations (56.6\,kJ/mol) by \citet{frolov_isomer_2005}, while the MP2 prediction is very close to the HEAT345(Q) value (56.3\,kJ/mol).  This small difference is likely fortuitous, owing to  a cancellation of errors, as the same agreement is not observed for $c$-\ce{H2C2S}. The HEAT345(Q) calculations predict that $c$-\ce{H2C2S} is much more stable (132.1\,kJ/mol) either compared to the MP2 (145.3\,kJ/mol) or B3LYP (155.2\,kJ/mol) predictions\cite{frolov_isomer_2005}. We attribute this stability to the improved treatment of dynamic correlation by coupled-cluster methods, combined with much larger basis sets used in the present calculation.

\ce{H2CCS} and \ce{HCCSH} are observed with comparable intensity in our electrical discharge when \ce{H2S} is used as the source of sulfur.  Because \ce{H2CCS} is much more polar than \ce{HCCSH} ($\mu_a$= 1.02 vs.\ 0.13\,D)\cite{Georgiou_1979}, however, this implies that \ce{HCCSH} was roughly four times more abundant under these conditions, taking into account nuclear spin statistics and differences in the rotational partition function. When \ce{CS2} is used instead of \ce{H2S}, lines of \ce{H2CCS} are observed with similar intensity, while those of \ce{HCCSH} are no longer detectable. With respect to \ce{H2CCS} formation, these results suggest the importance of atomic sulfur.   Although speculative, this finding is consistent with two pathways suggested by Yamada et al.\cite{Yamada:2002gm}: the reaction H + HCCS, where HCCS is presumably formed by the reaction HCC + S; or CH + HCS.  Because one pathway conserves the \ce{C2} unit in molecule formation, while the other does not, analogous $^{13}$C isotopic studies to those performed here should prove highly informative in clarify the pathways that yield thioketene from either \ce{H2S} and \ce{CS2}.


\subsection{Prospects for detection of higher-energy isomers}
 
A combination of discharge sources and supersonic jets has been used extensively to study higher-energy isomers\cite{brunken_laboratory_2009,mccarthy_laboratory_2007,mccarthy_high-resolution_2015,lattanzi_two_2012}, and  \ce{HCCSH} is no exception. Under some experimental conditions, \ce{HCCSH} is produced much more efficiently than the ground state isomer \ce{H2CCS}, despite the much lower stability of the former (56\,kJ/mol). The implication of this and previous studies is that collisional cooling near the throat of the expansion is fast relative to the timescale for unimolecular isomerization.  The efficiency of ``trapping'' energetic isomers appears particularly high when isomerization barriers are substantial, as has previously been calculated for \ce{HCCSH} $\leftrightarrow$ \ce{H2CCS} interconversion (${\sim}$80\,kJ/mol; Ref.~\citenum{gosavi_abinitio_1983}).
These results are in sharp contrast to the mechanisms at play in cryogenic matrices. \cite{krantz_characterization_1981,krantz_matrix_1974} In previous studies in argon matrices following UV irradiation of thiadazoles, it was found that \ce{H2CCS} formed first, and subsequent isomerization produced  higher-energy species such as \ce{HCCSH} and c-\ce{H2C2S}. At sufficiently long time scales however, \ce{HCCSH} is thought to rapidly tautomerize  to \ce{H2CCS}\cite{schaumann_chemistry_1988}.   

The present work suggests gas-phase detection of thiirene $c$-\ce{H2C2S} should be feasible with our FT microwave spectrometer. Theoretical calculations\cite{Yamada:2002gm} also conclude that once formed, there are sizable isomerization barriers (of order 100\,kJ/mol) to either \ce{H2CCS} or $c$-\ce{H2CCS}.  Given i) the high abundance of \ce{HSCCH} that can be achieved; ii) the high accuracy with which the rotational spectrum of $c$-\ce{H2C2S}  can likely be predicted; and iii) its high polarity, it would be surprising ---perhaps even disappointing--- if the rotational spectrum of this small, elusive antiaromatic heterocycle is not eventually be found. 

\subsection*{Astronomical implications}

Because there is very strong evidence the \ce{SH + CCH} reaction is responsible for \ce{HCCSH} in our discharge nozzle, and because the HEAT345(Q) energetics confirm this reaction is highly exothermic (-477\,kJ/mol), it is conceivable this metastable isomer might form preferentially in the interstellar medium, especially so since both the SH and CCH radicals are widely abundant and widely distributed there.  If relevant, this mechanism should be efficient even in low temperature environments such as dark molecular clouds (${\sim}20$\,K) which may make \ce{HCCSH} a viable sink of sulfur content there, a particularly intriguing possibility since this element is known to heavily depleted in these regions \cite{Boogert:1997ys,Bilalbegovic:2015dr,MartinDomenech:2016je}.

With the spectroscopic constants listed in Table \ref{tab:rotconstants}, it is possible to predict  the astronomically most interesting lines over the entire range of interest to radio astronomers.  Of particular importance is the strong $b$-type lines, which can now be predicted to better than 0.1\,km~sec$^{-1}$ in terms of equivalent radio velocity up to 800\,GHz. Using the experimental data, we are performing preliminary interstellar searches based on ALMA observations towards star-forming regions, as well as archival and publicly available datasets from the Green Bank Telescope and \textit{Herschel} Space Telescope. The analysis on \ce{HCCSH} and \ce{H2CCS} (e.g. upper limits on column densities) will be detailed in a forthcoming paper.

\section{Conclusions}
A high-resolution study on the gas-phase rotational spectrum of \ce{HCCSH}, the second most stable isomer with the elemental formula [\ce{H2},\ce{C2},S] was carried out. By measuring a total of 100 pure rotational transitions using on a combination of Fourier-transform microwave and submillimeter-wave spectroscopies guided by high level \textit{ab initio} predictions,  its rotational spectrum has been characterized from 10 to 660\,GHz. Accurate spectroscopic parameters have been determined from a fit of the experimental frequencies to a standard asymmetric top Hamiltonian with up to sextic centrifugal distortion constants.
The identity of \ce{HCCSH} was confirmed by detecting several microwave rotational transitions of six rare isotopic species: \ce{DCCSH}, \ce{HCCSD}, \ce{HCC^{34}SH}, \ce{H^{13}CCSH}, \ce{HC^{13}CSH}, and \ce{H^{13}C^{13}CSH}.  The isotopic measurements were beneficial in two other ways: to determine the molecular structure of HCCSH, and to infer its formation pathway in the electrical discharge, which we deduce to be radical recombination of \ce{SH + C2H \rightarrow HCCSH}. This reaction is highly exothermic (-477\,kJ/mol), as determined with the HEAT345(Q) method. The relative energetics of the [\ce{H2},\ce{C2},S] isomers have also been determined to higher accuracy compared to previously published methods.

With accurate predictions for transition frequencies of HCCSH up to about 1 THz, astronomical searches for this molecule can now be undertaken with confidence. While \ce{HCCSH} is not the ground state isomer, considering it is preferentially formed via radical-radical recombination of two abundant interstellar species --- namely \ce{SH} and \ce{C2H} --- and that kinetics rather than thermodynamic considerations often prevail in interstellar chemistry, this isomer appears to be a good candidate for astronomical detection. Using the same laboratory techniques, detection of $c$-\ce{H2C2S} would appear promising.

\section{Acknowledgements}

We thank Edward Tong of the Submillimeter Array Receiver laboratory at the Smithsonian  Astrophysical Observatory for the loan of a high-power, high-frequency VDI AMC that was used for the 300\,GHz DR measurements. The work in Cambridge is supported by NSF AST-1615847 and the work in Orsay by the Programme National ``Physique et Chimie du Milieu Interstellaire'' (PCMI) of CNRS/INSU with INC/INP co-funded by CEA and CNES, and by ``Investissements d'Avenir'' LabEx PALM (ANR-10-LABX-0039-PALM).  Support for B.A.M. was provided by NASA through Hubble Fellowship grant \#HST-HF2-51396 awarded by the Space Telescope Science Institute, which is operated by the Association of Universities for Research in Astronomy, Inc., for NASA, under contract NAS5-6555. The National Radio Astronomy Observatory is a facility of the National Science Foundation operated under cooperative agreement by Associated Universities, Inc.  

\section{Supplementary material}
Tables of frequencies for observed transitions, spectroscopic parameters used in the structure determination, and additional theoretical data can be found in supplementary material.
\bibliography{bibliography.bib}

\end{document}